\documentclass[aps,pra,showpacs,twocolumn,amsmath,amssymb]{revtex4-1}
\usepackage{graphicx}
\usepackage{amsmath}
\usepackage{color}
\usepackage{dcolumn}
\usepackage{bm}
\usepackage{booktabs}
\usepackage{subfigure}

\newcommand{\fHe}{$^4$He$^*$}
\newcommand{\tHe}{$^3$He$^*$}
\newcommand{\He}{He$^*$}
\newcommand{\mjp}{\mbox{$m$=$+1$}}
\newcommand{\mjm}{\mbox{$m$=$-1$}}
\newcommand{\etal}{\textit{et al.}}
\newcommand{\quintet}{$^5\Sigma_g^+$}
\newcommand{\triplet}{$^3\Sigma_u^+$}
\newcommand{\singlet}{$^1\Sigma_g^+$}

\begin{document}

\title{Magnetic-field dependent trap loss of ultracold metastable helium}

\author{J.\,S.\,Borbely}
\author{R.\,{van Rooij}}
\author{S.\,Knoop}
\author{W.\,Vassen}
\affiliation{LaserLaB Vrije Universiteit, De Boelelaan 1081, 1081 HV Amsterdam, the Netherlands}

\date{\today}

\begin{abstract}
We have experimentally studied the magnetic-field dependence of the decay of a Bose-Einstein condensate of metastable $^4$He atoms confined in an optical dipole trap, for atoms in the \mjp~and \mjm~magnetic substates, and up to 450~G. Our measurements confirm long-standing calculations of the two-body loss rate coefficient that show an increase above 50~G. We demonstrate that for \mjm~atoms, decay is due to three-body recombination only, with a three-body loss rate coefficient of $6.5(0.4)_{\rm stat}(0.6)_{\rm sys}\times10^{-27}$~cm$^6$s$^{-1}$, which is interesting in the context of universal few-body theory. We have also searched for a recently-predicted $d$-wave Feshbach resonance, but did not observe it.
\end{abstract}


\maketitle

\section{Introduction\label{Introduction}}

The realization of a Bose-Einstein condensate (BEC)~\cite{robert2001bec,pereira2001bec} and a degenerate Fermi gas~\cite{mcnamara2006dgb} of helium in the metastable 2~$^3$S$_1$ state (\He, radiative lifetime of 8000~s) has opened interesting possibilities for research~\cite{vassen2011cat}. Prominent examples are measurements of higher-order coherence in atomic matter-waves, including direct comparison between the bosonic and fermionic Hanbury Brown-Twiss effect~\cite{jeltes2007cot}, direct measurement of third-order coherence~\cite{hodgman2011dmo}, and production of squeezed states by four-wave mixing in colliding BECs~\cite{jaskula2010spn}. These experiments take advantage of the 19.8~eV internal energy of \He~atoms, which allows for single atom detection with high spatial and temporal resolution by using micro-channel plate detectors. Also, ultracold trapped \He~allows for precise spectroscopy of very weak atomic transitions~\cite{rooij2011fmi}, of interest for fundamental tests of two-electron quantum electrodynamic theory.

The realization of ultracold and dense samples of \He~atoms is quite remarkable since the large internal energy allows for detrimental Penning (and associative) ionization loss processes due to collisions between two \He~atoms. The corresponding inelastic rate constant, $\sim 1\times10^{-10}$ cm$^3$s$^{-1}$~\cite{julienne1989cou,vassen2011cat}, would limit evaporative cooling of \He~atoms and inhibit the possibility of achieving BEC. However, in a gas of spin-polarized atoms, Penning ionization is forbidden by spin conservation and leads to a suppression of inelastic collision rates~\cite{julienne1989cou}. Shlyapnikov \etal~\cite{shlyapnikov1994dka} considered the magnetically trappable 2~$^3$S$_1$, \mjp~state (where $m$ is the magnetic quantum number) of \fHe~and found Penning ionization to be suppressed by four orders of magnitude, indicating the possibility of BEC for \fHe. The strongest inelastic two-body processes for \mjp~were found to be spin-relaxation (SR) and relaxation-induced Penning ionization (RIPI), both induced by the spin-dipole interaction. At zero magnetic field a rate constant of 2$\times10^{-14}$ cm$^3$s$^{-1}$, dominated by RIPI, was calculated~\cite{shlyapnikov1994dka}, which was confirmed by other calculations~\cite{fedichev1996idp,venturi1999ccc,leo2001uco}. Several experiments measured losses in magnetic traps in agreement with this loss rate~\cite{herschbach2000sop,pereira2001bec,sirjean2002iri,seidelin2004gte,tychkov2006mtb}. For the high densities typically present in a BEC, trap loss caused by three-body recombination competes with two-body loss. A three-body loss rate of 2$\times10^{-27}$ cm$^6$s$^{-1}$ was calculated by Fedichev \etal~\cite{fedichev1996tbr}.

Recently, trapping of \fHe~\cite{partridge2010bec,dall2011ooa,rooij2011fmi} and \tHe~\cite{rooij2011fmi} in optical dipole traps (ODT) has been demonstrated. This has opened new possibilities over magnetic trapping. Most notably, it allows for trapping of the $m$=0 and \mjm~spin states~\cite{partridge2010bec} and the application of Feshbach resonances to control the scattering properties by a magnetic field. Two-body loss rates of low density $m=0,\pm1$ spin mixtures in an ODT for a small, fixed magnetic field have recently been measured~\cite{partridge2010bec}. This study confirmed strong Penning ionization (loss rates on the order of 10$^{-10}$ cm$^3$s$^{-1}$) for those spin mixtures. In this paper we present trap loss measurements in \fHe~for single-spin \mjp~and \mjm~clouds, which are expected to show suppression of Penning ionization, for fields up to 450~G. In particular, we have investigated the prediction of a strong magnetic-field dependence of the two-body loss rate for atoms in the \mjp~state~\cite{shlyapnikov1994dka,fedichev1996idp,venturi1999ccc}, which had not yet been experimentally tested. We have also searched for a $d$-wave Feshbach resonance, which was recently predicted~\cite{goosen2010fri}.

This paper is organized as follows. In Sec.~\ref{Heproperties} we briefly discuss the collisional properties of spin-polarized, ultracold \fHe~with a particular interest to the dependence with magnetic field. In Sec.~\ref{loss} we discuss the time evolution of a trapped gas. In Sec.~\ref{Exp} we outline our experimental setup. In Sec.~\ref{Results} we present our experimental results and compare to theoretical data from the literature. Finally, in Sec.~\ref{conclusion} we conclude and give an outlook for using our measured three-body loss rate coefficient in the context of universal theory and for magnetic-field dependent trap losses in ultracold \tHe-\fHe~mixtures, where an 800-G broad Feshbach resonance is predicted~\cite{goosen2010fri}.

\section{Collisional properties of spin-polarized ultracold \fHe}\label{Heproperties}

The collisional properties of an ultracold atomic gas, dominated by $s$-wave collisions, are intimately linked to underlying two-body potentials. The electron spin of a \He~atom with $s=1$ gives rise to three distinct Born-Oppenheimer (BO) potentials: singlet ($S=0$)~\singlet, triplet ($S=1$)~\triplet~and quintet ($S=2$)~\quintet, where the total electronic spin is given as $\vec{S}=\vec{s}_1+\vec{s}_2$. Since \fHe~has no nuclear spin the total atomic spin is equal to the electron spin, $s$, with projection $m$. The Hamiltonian of the BO potentials is spherically symmetric and therefore conserves the total electron spin projection, $M_S$. For the interaction between two atoms in either the \mjp~or \mjm~state the total spin projection is $|M_S|=2$, and therefore scattering is only given by the \quintet~potential. The energy of the least bound state of the \quintet~potential was measured to be $h\times91.35(6)$~MHz~\cite{moal2006ado}, from which a precise quintet scattering length of 142.0(0.1)~$a_0$ (where $a_0$ is the Bohr radius) has been derived. The absence of hyperfine coupling between the different BO potentials excludes the possibility of $s$-wave Feshbach resonances with spin-stretched \fHe~atoms.

Penning ionization (PI) and associative ionization (AI) play an important role in ultracold \He~gases. For an unpolarized sample, the following two-body loss processes limit the stability of the trapped sample:
\begin{equation}
\label{PI}
{\rm He}^*+{\rm He}^* \rightarrow \Biggl\{ \begin{array}{lr}{\rm He}+{\rm He}^++{\rm e}^- & \qquad{\rm (PI)}\\ {\rm He}_2^++{\rm e}^- & \qquad{\rm (AI)} \end{array}
\end{equation}
In the following, we will refer to both processes in Eq.~\ref{PI} as PI.

PI is spin-forbidden for scattering in the \quintet~potential, since the total spin of the final PI state cannot exceed 1. Therefore, a spin-polarized sample, i.e., atoms prepared in either the \mjp~or \mjm~state, is stable. However, weak higher-order interactions can couple the different BO potentials which will induce loss processes. Shlyapnikov \etal~\cite{shlyapnikov1994dka} identified the spin-dipole interaction as the most important higher-order interaction that leads to a weak coupling of the $S=2$, $M_S=2$ state to the $S=2$, $M_S=0,1$ and $S=0$, $M_S=0$ states. For scattering between identical particles in absence of nuclear spin, the condition $S+\ell$=even (where $\ell$ is the total angular momentum) is required, which excludes coupling to the triplet \triplet~potential. For spin-dipole interactions only $M_S+M_{\ell}$ is conserved, where $M_\ell$ is the projection of $\ell$, and $\Delta \ell=0,2$ so that the final channel after the spin-dipole interaction is characterized by $\ell=2$, i.e., $d$-waves. Furthermore, the spin-dipole interaction allows for coupling between the quintet scattering state and the singlet molecular state with $\ell=2$ and therefore $d$-wave Feshbach resonances are possible. Ref.~\cite{goosen2010fri} predicted the existence of a $d$-wave Feshbach resonance, either in a pure \mjm~sample below 470~G or in a pure \mjp~sample below 90~G, by varying the singlet potential within the theoretical bounds. In the latter case the molecular state would be a shape resonance.

The spin-dipole interaction induces two inelastic, two-body processes: SR, where the energy gain or loss is determined by the Zeeman energy and RIPI which is due to coupling of the $S=2$, $M_S=2$ state to the $S=0$, $M_S=0$ state, which is a strongly Penning ionizing state. At zero magnetic field the loss rate due to both SR and RIPI is independent of $m$ and therefore is equal for \mjp~and \mjm~states. However, at magnetic fields for which $2\mu_B B\gg k_B T$, both processes are energetically not allowed for \mjm~states. This has the consequence that in a 1~$\mu$K cloud of \mjm~\He~atoms SR and RIPI can be neglected for magnetic fields larger than 10~mG. The much weaker direct dipole-exchange mechanism and spin-orbit coupling might still be possible, but are estimated to have rates $\lesssim 10^{-16}$ cm$^3$s$^{-1}$~\cite{shlyapnikov1994dka,fedichev1996idp} and therefore do not play any role here. For \mjp~atoms trapped in a magnetic trap, SR leads to a transfer to untrapped $m$=0 and \mjm~states. In an optical trap SR induces two different loss mechanisms. For sufficiently high magnetic fields, i.e., $B>0.1$~G for a trap depth of 10~$\mu$K, the high gain in kinetic energy of the $m$=0 and \mjm~ reaction products induces instant trap loss of those atoms. For $B<0.1$~G, the large PI loss rate constant for collisions between \mjp~and \mjm~atoms and/or between $m$=0 atoms will also remove those atoms from the trap.

\begin{figure}[b]
\includegraphics[width=8.5cm]{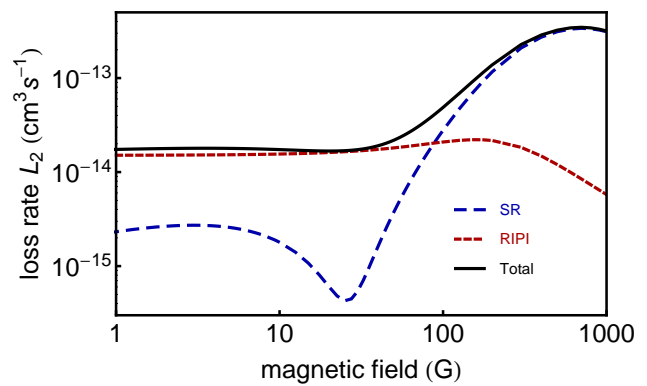}
\caption{(Color online) Two-body loss rate coefficient $L_2$ of \fHe~in the \mjp~state at 1~nK (obtained from close-coupling calculations by Venturi \etal~\cite{venturi1999ccc}), showing the contribution from relaxation-induced Penning ionization (RIPI) and from spin-relaxation (SR).
\label{venturi}}
\end{figure}

The magnetic-field dependence of SR and RIPI rate constants for \mjp~was investigated in several papers, using perturbative methods~\cite{shlyapnikov1994dka,fedichev1996idp} and close-coupling calculations~\cite{venturi1999ccc}, all showing similar behavior that for small magnetic fields RIPI dominates with a rate of $\sim2\times10^{-14}$~cm$^3$s$^{-1}$ whereas for large magnetic fields SR becomes the dominant loss mechanism with a maximum rate of $\sim3\times10^{-13}$~cm$^3$s$^{-1}$ at a field of 700~G (see Fig.~\ref{venturi}). Around 100~G a crossover between the two loss processes occurs due to the strong magnetic-field dependence of SR~\cite{fedichev1996idp,venturi1999ccc}. An increase in trap loss is therefore expected for $B>50$~G. The calculation for SR is very sensitive to the \quintet~potential and to a lesser degree to the \singlet~potential. The calculations of Refs.~\cite{shlyapnikov1994dka,fedichev1996idp,venturi1999ccc} were based on the \quintet~potential of St\"arck and Meyer \cite{stark1994lri} and on the \singlet~potential of M\"uller \etal~\cite{muller1991eat}, which was modified to have the same long-range potential as the \quintet~potential. More accurate calculations of the \quintet~potential have been performed since then~\cite{przybytek2005bft}, which, within the theoretical bounds, are in good agreement with the latest experimental value of the quintet scattering length~\cite{moal2006ado}. The difference between the rate coefficients obtained by these different \quintet~potentials is, at most, a factor of two~\cite{simonet2011PhD}.

For large enough densities, trap loss caused by three-body recombination (TBR), described by
\begin{eqnarray}
\label{TBR}
{\rm He}^*+{\rm He}^*+{\rm He}^* \rightarrow {\rm He}^*_2 + {\rm He}^* \qquad{\rm (TBR)}
\end{eqnarray}
will compete with two-body loss processes. ${\rm He}^*_2$ in Eq.~\ref{TBR} will undergo fast PI. The energy gain in TBR is given by the binding energy of the least bound state of the \quintet~potential, which in temperature units is 4~mK, leading to loss in both magnetic and optical dipole traps. TBR depends strongly on the scattering length, $a$. In particular, when $a$ is much larger than the van der Waals length, $r_{\rm vdW}$, the rate coefficient for TBR, $L_3$, is given by
\begin{equation}
\label{L3TBR}
L_3=3C(a)\frac{\hbar}{m}a^4,
\end{equation}
where, for $a>0$, $C(a)$ is an oscillating function between 0 and 70 with an unknown phase~\cite{braaten2006uif} and assuming that three atoms are lost from the trap~\cite{weber2003tbr}. For \fHe, $a/r_{\rm vdW}\approx4$~\footnote{The van der Waals length $r_{\text{vdW}}=\frac{1}{2}(m C_6/\hbar^2)^{1/4}$=34$a_0$, with $C_6$=3276.680 a.\,u.\ from~\cite{przybytek2005bft}}, and universal few-body physics related to a large scattering length~\cite{braaten2006uif} can be expected. Since the scattering length for \mjp~and \mjm~atoms is equal and magnetic-field independent, so is the three-body loss rate coefficient.

\section{Trap loss equation}\label{loss}

In order to study the different loss processes one has to monitor the time-evolution of the density of a trapped atomic gas, which can be described as:
\begin{equation}
\label{densityevolution}
\dot{n}=-n/\tau-\kappa_2 L_2 n^2-\kappa_3 L_3 n^3.
\end{equation}
The first term takes into account one-body loss, mainly due to collisions with background gas, which causes exponential decay with a time constant $\tau$. $L_2$ and $L_3$ are the rate coefficients for two- and three-body loss, respectively, and are defined such that they explicitly include the loss of two and three atoms per loss event. The constants in front of $L_2$ and $L_3$ are $\kappa_2=1/2!$ and $\kappa_3=1/3!$ for a BEC (where we neglect quantum depletion~\cite{sirjean2002iri}), while $\kappa_2=\kappa_3=1$ for a thermal gas~\cite{kagan1985}.

Since we measure the atom number, $N$, we have to integrate Eq.~\ref{densityevolution}, which for a BEC in the Thomas-Fermi regime gives
\begin{equation}
\label{atomnumberevolution}
\dot{N}=-N/\tau-\kappa_2 b_2 L_2 N^{7/5}-\kappa_3 b_3 L_3 N^{9/5},
\end{equation}
where $b_2=1/(210\pi)a^{-2}a_{\rm ho}^{-1}\left(15a/a_{\rm ho}\right)^{7/5}$ and \linebreak $b_3=1/(2520\pi^2)(aa_{\rm ho})^{-3}\left(15a/a_{\rm ho}\right)^{9/5}$, with harmonic oscillator length $a_{\rm ho}=\sqrt{\hbar/(m\bar{\omega})}$ and the geometric mean of the trap frequencies $\bar{\omega}=2\pi(\nu_{\rm ax} \nu_{\rm rad}^2)^{1/3}$~\cite{pitaevskii2003book}. Analytical solutions of Eq.~\ref{atomnumberevolution} can be found for pure two- or three-body loss, but in general one has to solve Eq.~\ref{atomnumberevolution} numerically.

\section{Experimental setup}\label{Exp}

\begin{figure}[t]
\includegraphics[width=8.5cm]{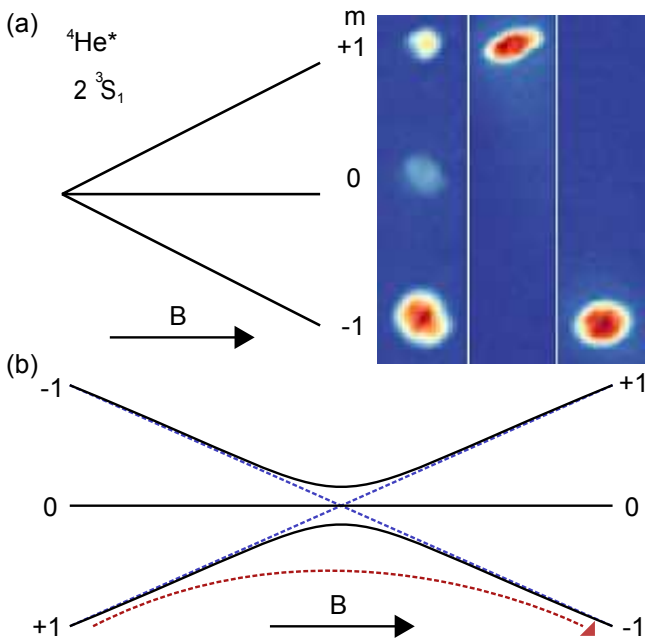}
\caption{(Color online) (a) Zeeman diagram of \fHe~and absorption images that show the population of the magnetic substates of the 2 $^3$S$_1$ manifold after a Stern-Gerlach type experiment for (left to right) a mixture of $m$=0,$\pm$1 atoms, spin-polarized \mjp~atoms and spin-polarized \mjm~atoms. (b) Dressed picture of \fHe~in an RF field, leading to an avoided crossing at $B=h\nu_{RF}/(2\mu_B)$. An adiabatic passage over the avoided crossing leads to transfer from \mjp~to \mjm~and vice versa.
\label{transfer}}
\end{figure}

Our experimental setup and cooling procedure has been outlined earlier~\cite{tychkov2006mtb,rooij2011fmi}. In short, we use a liquid-nitrogen cooled dc-discharge source to produce a beam of metastable helium atoms that is collimated, slowed and loaded into a magneto-optical trap in 2~s. The atomic gas is optically pumped into the \mjp~state, after which it is loaded into a cloverleaf magnetic trap. After 2.5~s of 1D-Doppler cooling and 5~s of forced evaporative RF cooling, BEC is realized. We transfer up to 10$^6$ atoms into a crossed ODT at 1557~nm, which is formed by two beams that are focused to a waist of 85~$\mu$m at the intersection and cross under an angle of 19$^\circ$ in the horizontal plane~\cite{rooij2011fmi}. The power used for our ODT is between 100 and 500~mW. A small, uniform magnetic field is applied to ensure that atoms stay in the \mjp~state after they have been transferred into the ODT. To prepare a spin-polarized sample in the \mjm~state, a small magnetic field sweep from 1 to 2~G in 50~ms is applied while the atoms are in an RF field at a fixed frequency to transfer atoms from the \mjp~state to the \mjm~state with nearly 100\% efficiency (see Fig.~\ref{transfer}).

Once atoms are trapped in the ODT, the axial compensation coils of the cloverleaf magnetic trap are used to create the required magnetic field. The magnetic field is calibrated by performing spin flips between \mjp~and \mjm~as described above, recording the RF resonance frequency at different currents applied through the coils. Because the coils are not in a geometrically ideal Helmholtz configuration, the field creates an anti-trapping potential for \mjm~and a trapping potential for \mjp, with a curvature of about 0.1~G/cm$^2$ for a field of 1~G. This curvature primarily affects the axial trap frequency. The decrease (\mjm) and increase (\mjp) of the axial trap frequency is 19\% at 450~G for the loss rate measurements presented in this paper and is corrected for in the analysis. Furthermore, the curvature effectively leads to a decrease of the trap depth for atoms in the \mjm~state, so that for a particular ODT power there is a maximum magnetic field beyond which all Bose-condensed \mjm~atoms escape from the trap. The inhomogeneity of the magnetic field across the BEC in the ODT is on the order of 1~mG and plays no role in this study.

The number of trapped \fHe~atoms is measured by turning off the ODT, which causes the atoms to fall and be detected by a micro-channel plate (MCP) detector, which is located 17~cm below the trap center and gives rise to a time-of-flight of approximately 186~ms. From a bimodal fit to the MCP signal, the BEC and thermal fraction are extracted as well as the temperature and the chemical potential, $\mu$, of the trapped gas~\cite{rooij2011fmi,partridge2010bec}. We use absorption imaging for setting up the transfer scheme between \mjp~and \mjm~states by using Stern-Gerlach separation, as shown in Fig.~\ref{transfer}(a), and to measure trap frequencies by recording induced trap oscillations.

\begin{figure}[b]
\includegraphics[width=8.5cm]{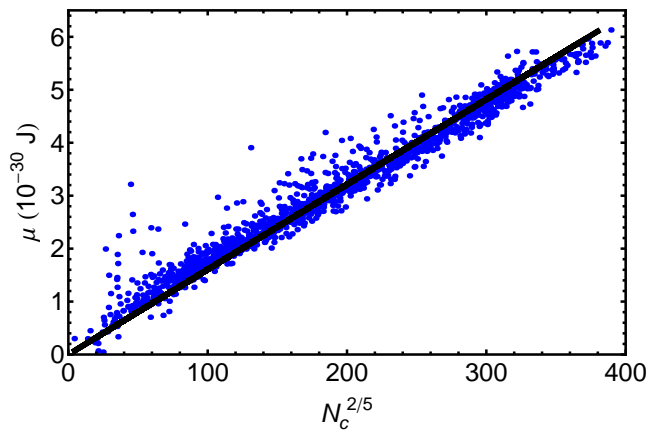}
\caption{(Color online) Plot of the chemical potential, $\mu$, versus the number of condensed atoms to the power of $2/5$. The MCP signal gives a relative measure for the number of condensed atoms, $V_{\rm MCP}$, and therefore this signal is corrected by a factor $\alpha$ to represent the number of condensed atoms, $N_c=\alpha V_{\rm MCP}$. The value of $\alpha$ is determined such that the MCP signal (circles), for all data acquired, has a slope that equals the theoretical slope of $\frac{1}{2}(15a\bar{\omega}^{3}\hbar^{2}M^{1/2})^{2/5}$ (line).
\label{muvsNc}}
\end{figure}

The BEC part of the signal from the MCP detector ($V_{\rm MCP}$) relates to the number of condensed atoms as $N_c=\alpha V_{\rm MCP}$, where $\alpha$ is a conversion factor dependent on several factors, such as the applied potential difference across the MCP detector and on its detection efficiency. In the Thomas-Fermi limit, the relation between $\mu$ and the number of condensed atoms is $2\mu=(15a\bar{\omega}^{3}\hbar^{2}M^{1/2})^{2/5}N_c^{2/5}$~\cite{pitaevskii2003book}, where $M$ is the mass of a helium atom, and is used to determine $\alpha$ (see Fig.~\ref{muvsNc}). Since the scattering length is known~\cite{moal2006ado} and the value for the average trap frequency, $\bar{\omega}=2\pi(\nu_{\rm ax} \nu_{\rm rad}^2)^{1/3}$, where $\nu_{\rm ax} = 55.3(0.3)$~Hz and $\nu_{\rm rad} =363.4(2.1)$~Hz, was measured, the theoretical slope of a $\mu$ versus $N_c^{2/5}$ plot is known. A value of $\alpha$ was determined, for the entire data set, such that the slope of $\mu$ vs $(\alpha V_{\rm MCP})^{2/5}$ equals the theoretical slope.

The determination of $\alpha$ for atoms in an ODT is much more reliable than in a magnetic trap. In the case of atoms in a magnetic trap, the velocity distribution may be distorted during trap switch-off as a result of magnetic-field gradients on the initial expansion of the atomic cloud. These gradients can lead to an overestimation of $\mu$~\cite{tychkov2006mtb}. A ballistic expansion from the ODT, however, is not hindered by these effects.

The experimental procedure to study magnetic-field dependent trap loss is as follows. Atoms are confined in the ODT and are in the \mjp~state. Next, a small magnetic field of 0.5~G is applied to ensure a quantization axis. Atoms are then transferred to the \mjm~state using the spin-flip procedure illustrated in Fig.~\ref{transfer}. The reason for performing this initial spin-flip is that absorption imaging showed that, unlike \mjm~atoms, hot atoms in the \mjp~state would remain trapped in the wings of the crossed ODT due to the additional trapping force resulting from the residual magnetic-field curvature. After 500~ms of rethermalization time, either another spin-flip procedure is performed in which case trap loss experiments for \mjp~atoms are investigated, or, atoms remain in the \mjm~state. The magnetic field is then ramped up to a certain value in 100~ms, the atoms remain in this field for a variable time (ranging from 10~ms to 50~s), after which the magnetic field is turned off. Finally, the small 0.5~G field and the ODT are both turned off and the number of metastables is monitored by the MCP detector.

\section{Results}\label{Results}
\subsection{Two- and three-body loss}

\begin{figure}
\includegraphics[width=8.5cm]{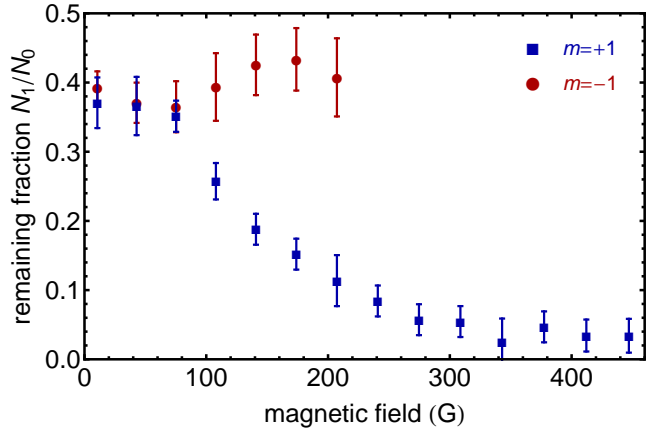}
\caption{(Color online) Fraction of atoms, in the \mjp~and \mjm~states, remaining in the ODT as a function of magnetic field. $N_0$ and $N_1$ represent the number of condensed atoms remaining after exposed to a magnetic field for 10~ms and 2~s, respectively. Two-body loss processes are energetically not allowed for atoms in the \mjm~state and therefore the total loss rate is dominated by the three-body loss rate which is independent of magnetic field.
\label{fraction}}
\end{figure}

We have performed two types of measurements: (1) we compare the total magnetic-field dependent loss rate for atoms in the \mjp~state versus the \mjm~state, and (2) we monitor the lifetime of the BEC for various magnetic fields for atoms in both the \mjp~state and the \mjm~state. The first measurement illustrates the differences between atoms in the \mjp~and \mjm~states, in particular, that \mjm~atoms have no magnetic-field dependent loss processes, as shown in Fig.~\ref{fraction}. The experimental procedure is to measure the number of atoms remaining for a fixed hold time of either 10~ms (representing the initial atom number $N_0$) or 2~s (representing the final atom number $N_1$), in the presence of various magnetic fields between 10 and 450~G for atoms in both the \mjp~and \mjm~states. The remaining fraction, $N_1/N_0$, gives insight into the magnetic-field dependence of the loss processes. At fields $<$100~G the total loss for \mjp~and \mjm~samples are equal since three-body loss is the dominating mechanism and the two-body loss rate remains small at these fields. At approximately 100~G SR begins to contribute significantly to the total loss and a clear difference between the \mjp~and \mjm~states becomes evident. Atoms in the \mjm~state cannot undergo two-body loss processes because it is energetically not allowed and since three-body loss process are magnetic-field independent the remaining fraction for an \mjm~sample remains constant as a function of magnetic field. Due to the curvature of the magnetic field, as discussed in Sec.~\ref{Exp}, there is a maximum magnetic field for which the depth of the ODT ($\sim$1~$\mu$K) is sufficient to confine \mjm~atoms, which was 210~G in this case.

The second measurement is to monitor the time-evolution of the number of condensed atoms and solve Eq.~\ref{atomnumberevolution} to determine $\tau$, and $L_2$ and $L_3$ as a function of magnetic field. Since $\tau$ characterizes background collisions it is independent of magnetic field and insensitive to whether atoms are in the \mjp~or \mjm~state. By looking at trap loss for atoms in a 10~G field at long (20 to 40~s) hold times a value of $\tau$$\sim$25~s was determined. As it can be difficult to distinguish between two- and three-body loss, we first extract $L_3$ by looking at trap loss for atoms in the \mjm~state (for which two-body loss is absent) and use this value to determine $L_2$ as a function of magnetic field for atoms in the \mjp~state. This is valid since three-body loss rates are equal for both spin states. We have shown that three-body loss is indeed independent of magnetic field (see Fig.~\ref{fraction} for \mjm~atoms) and we have determined a value of $L_3=6.5(0.4)_{\rm stat}(0.6)_{\rm sys}\times10^{-27}$~cm$^6$s$^{-1}$. The systematic uncertainty is due to the propagation in uncertainties of $a$, $\tau$, $\nu_{\rm ax}$, $\nu_{\rm rad}$ and the conversion factor $\alpha$. The present value of $L_3$ is in fairly good agreement with previous experimental results, but more accurate. Previously, our group has measured the three-body loss rate to be 9(3)$\times10^{-27}$~cm$^6$s$^{-1}$~\cite{tychkov2006mtb,vassen2011cat}, the experiment of Seidelin \etal~\cite{seidelin2004gte} determined 0.8$^{+1.4}_{-0.5}\times10^{-27}$~cm$^6$s$^{-1}$ (corrected~\cite{sirjean2002iri} for the current value of $a$), and an upper limit of 1.7(1)$\times10^{-26}$~cm$^6$s$^{-1}$ was given by Pereira Dos Santos \etal~\cite{pereira2001bec}. Our value is larger than the only published theoretical value of $L_3$ of $2\times10^{-27}$~cm$^6$s$^{-1}$~\cite{fedichev1996idp,fedichev1996tbr}, in which a scattering length independent prefactor in front of the $a^4$ scaling was assumed (see Eq.~\ref{L3TBR}), with $C=3.9$, and a larger scattering length of 190$a_0$ was used.

\begin{figure}[b]
\includegraphics[width=8.5cm]{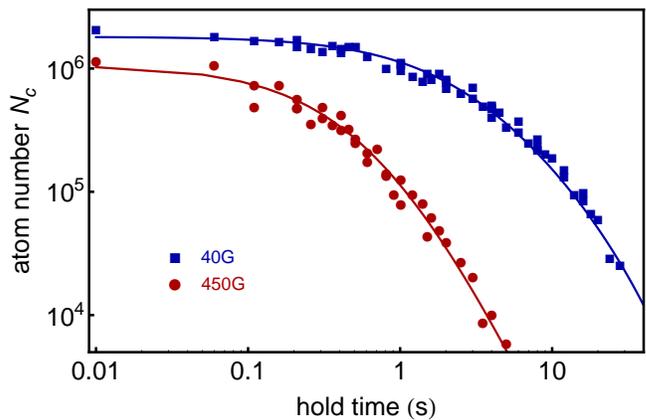}
\caption{(Color online) Log-log plot showing the number of condensed \mjp~atoms versus hold time at 40 and 450~G. The solid lines are numerical evaluations of Eq.~\ref{atomnumberevolution}, fitted to the data, with a fixed value of $\tau$=25~s and $L_3$=6.5$\times10^{-27}$~cm$^6$s$^{-1}$, and a variable $L_2$ rate coefficient.
\label{L2lifetime}}
\end{figure}

Having determined the one-body lifetime $\tau$ and the three-body loss rate $L_3$ we are now in the position to extract the two-body loss rates $L_2$ by numerically integrating Eq.~\ref{atomnumberevolution} for \mjp~atoms. Fig.~\ref{L2lifetime} shows the lifetime of a trapped sample of \mjp~atoms in a low (squares) and a high (circles) magnetic field. For \mjp~atoms in low magnetic fields, theory predicts that two-body losses are small and remain relatively constant up to 50~G (see Fig.~\ref{venturi}). We indeed observe that for small magnetic fields ($B<75$~G) our \mjp~data are fully dominated by three-body loss. For large magnetic fields ($B>75$~G) the inclusion of two-body loss is required to fit the data.

\begin{figure}[t]
\includegraphics[width=8.5cm]{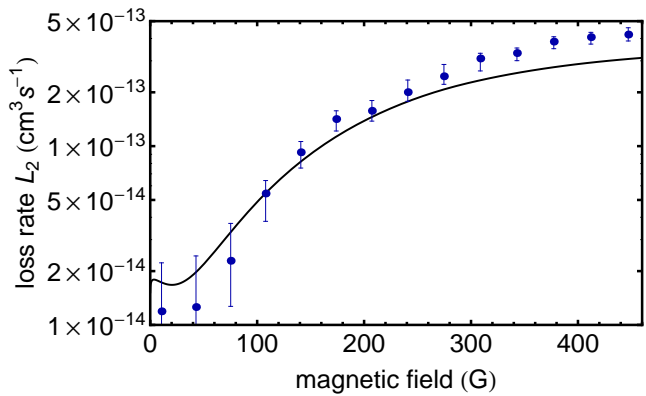}
\caption{(Color online) Comparison between experiment (points) and theory~\cite{venturi1999ccc} (line) for the two-body loss rate coefficient $L_2$ as a function of magnetic field. The uncertainties in the value of $L_2$ for $B<50$~G have a minimum value of 0 and are not shown in the plot.
\label{l2vsb}}
\end{figure}

We show our experimental results for the two-body loss rate $L_2$ as a function of magnetic field in Fig.~\ref{l2vsb}, together with theory~\cite{venturi1999ccc}.
Our low field data are in good agreement with previous experimental results at zero magnetic field: 2(1)$\times10^{-14}$~cm$^3$s$^{-1}$~\cite{tychkov2006mtb,vassen2011cat}, 0.4$^{+0.7}_{-0.3}\times10^{-14}$~cm$^3$s$^{-1}$~\cite{seidelin2004gte} (corrected~\cite{sirjean2002iri} for the current value of $a$), and an upper limit of 8.4(1.2)$\times10^{-14}$~cm$^3$s$^{-1}$~\cite{pereira2001bec}. Our result shows good agreement between experiment and theory at magnetic fields up to 250~G but deviates, up to approximately a factor of 1.3, at higher magnetic fields. The theoretical evaluation is strongly dependent on the \quintet~potential. The long-range part of the potential is well known; however, the short-range part of the potential is not well known and therefore our data can be used to correct the short-range part of the potential~\cite{shlyapnikov1994dka,simonet2011PhD}.

\subsection{Feshbach resonance}
We have searched for the narrow $d$-wave Feshbach resonance, caused by a $\ell$=2 singlet molecular state, with a predicted width of 20~mG~\cite{goosen2010fri}. In general, at a Feshbach resonance loss processes are enhanced, leading to loss resonances as a function of magnetic field. The 20~mG width of this resonance is broad enough to be observed as resonant loss in our BEC (see Ref.~\cite{knoop2011fsa} where even much narrower resonances down to $2\times10^{-4}$~mG were observed, although with a 10$\times$ higher density).

We have scanned magnetic fields up to 560~G for \mjm~(using both the compensation and pinch coils to reach fields above 450~G, and a much deeper ODT trap) and 120~G for \mjp, fully covering the predicted range.  We have used small magnetic field sweeps of 1-2~G in a few seconds, but did not observe any resonant enhanced loss. One possible explanation is that the finite lifetime of the singlet molecular state, due to Penning ionization, leads to a broadening of the Feshbach resonance~\cite{goosen2010fri}. We expect the lifetime of the singlet molecular state to be much shorter than that of the quintet molecular state, which is 1~$\mu$s~\cite{moal2006ado}.

\section{Conclusion and Outlook}\label{conclusion}

We have investigated the stability of ultracold spin-polarized gases of \fHe~in both \mjp~and \mjm~states. We have experimentally confirmed the long-standing theoretical prediction of a magnetic-field dependent two-body loss rate, which limits the stability of atoms in the \mjp~state at magnetic fields above 100~G. For \mjm~atoms two-body loss is energetically not possible (when $2\mu_B B\gg k_B T$), and therefore only three-body loss limits the lifetime of a dense sample, which is independent of magnetic field. We have also searched for a $d$-wave Feshbach resonance but did not observe one within the range of magnetic fields predicted by theory. We expect the short lifetime of the corresponding molecular state of the \singlet~potential to be responsible for smearing out of the resonance.

Our measured $L_3$ coefficient is interesting in the context of universal few-body physics~\cite{braaten2006uif}, as the interactions in a spin-polarized \fHe~gas are governed by a large two-body scattering length with $a/r_{\rm vdW}\approx4$. We note that recent theoretical few-body studies have focussed on ground state $^4$He and alkali atoms with tunable scattering length around a Feshbach resonance. Even though spin-polarized \fHe~atoms do not provide a tunable scattering length, it does provide an interesting benchmark system because of very accurate knowledge of the two-body potential~\cite{przybytek2005bft} and can be used to study effects beyond universal theory (i.\,e.\,, Refs.~\cite{hammer2007erc,platter2009rct,ji2010bui}).

Future experiments will advance in the direction of ultracold \tHe-\fHe~mixtures~\cite{mcnamara2006dgb}. \tHe~has nuclear spin, but because of its inverted hyperfine splitting, the lowest spin state is spin-stretched. Therefore, a mixture of \tHe-\fHe~prepared in the lowest spin channel is also stable against Penning ionization as well as SR and RIPI which are energetically not allowed. This stability provides an ideal starting point to prepare an ultracold mixture in the dipole trap at large magnetic fields in order to explore the recently-predicted 800-G broad interspecies Feshbach resonance~\cite{goosen2010fri}.

\begin{acknowledgments}

We thank Maikel Goosen, Servaas Kokkelmans and Juliette Simonet for helpful discussions, and Vanessa Venturi for providing us published theoretical data. We acknowledge Jacques Bouma for technical support. This work was financially supported by the Dutch Foundation for Fundamental Research on Matter (FOM). S.\,K.\ acknowledges financial support from the Netherlands Organization for Scientific Research (NWO) via a VIDI grant.

\end{acknowledgments}


\begin{thebibliography}{35}%
\makeatletter
\providecommand \@ifxundefined [1]{%
 \@ifx{#1\undefined}
}%
\providecommand \@ifnum [1]{%
 \ifnum #1\expandafter \@firstoftwo
 \else \expandafter \@secondoftwo
 \fi
}%
\providecommand \@ifx [1]{%
 \ifx #1\expandafter \@firstoftwo
 \else \expandafter \@secondoftwo
 \fi
}%
\providecommand \natexlab [1]{#1}%
\providecommand \enquote  [1]{``#1''}%
\providecommand \bibnamefont  [1]{#1}%
\providecommand \bibfnamefont [1]{#1}%
\providecommand \citenamefont [1]{#1}%
\providecommand \href@noop [0]{\@secondoftwo}%
\providecommand \href [0]{\begingroup \@sanitize@url \@href}%
\providecommand \@href[1]{\@@startlink{#1}\@@href}%
\providecommand \@@href[1]{\endgroup#1\@@endlink}%
\providecommand \@sanitize@url [0]{\catcode `\\12\catcode `\$12\catcode
  `\&12\catcode `\#12\catcode `\^12\catcode `\_12\catcode `\%12\relax}%
\providecommand \@@startlink[1]{}%
\providecommand \@@endlink[0]{}%
\providecommand \url  [0]{\begingroup\@sanitize@url \@url }%
\providecommand \@url [1]{\endgroup\@href {#1}{\urlprefix }}%
\providecommand \urlprefix  [0]{URL }%
\providecommand \Eprint [0]{\href }%
\providecommand \doibase [0]{http://dx.doi.org/}%
\providecommand \selectlanguage [0]{\@gobble}%
\providecommand \bibinfo  [0]{\@secondoftwo}%
\providecommand \bibfield  [0]{\@secondoftwo}%
\providecommand \translation [1]{[#1]}%
\providecommand \BibitemOpen [0]{}%
\providecommand \bibitemStop [0]{}%
\providecommand \bibitemNoStop [0]{.\EOS\space}%
\providecommand \EOS [0]{\spacefactor3000\relax}%
\providecommand \BibitemShut  [1]{\csname bibitem#1\endcsname}%
\let\auto@bib@innerbib\@empty
\bibitem [{\citenamefont {Robert}\ \emph {et~al.}(2001)\citenamefont {Robert},
  \citenamefont {Sirjean}, \citenamefont {Browaeys}, \citenamefont {Poupard},
  \citenamefont {Nowak}, \citenamefont {Boiron}, \citenamefont {Westbrook},\
  and\ \citenamefont {Aspect}}]{robert2001bec}%
  \BibitemOpen
  \bibfield  {author} {\bibinfo {author} {\bibfnamefont {A.}~\bibnamefont
  {Robert}}, \bibinfo {author} {\bibfnamefont {O.}~\bibnamefont {Sirjean}},
  \bibinfo {author} {\bibfnamefont {A.}~\bibnamefont {Browaeys}}, \bibinfo
  {author} {\bibfnamefont {J.}~\bibnamefont {Poupard}}, \bibinfo {author}
  {\bibfnamefont {S.}~\bibnamefont {Nowak}}, \bibinfo {author} {\bibfnamefont
  {D.}~\bibnamefont {Boiron}}, \bibinfo {author} {\bibfnamefont {C.~I.}\
  \bibnamefont {Westbrook}}, \ and\ \bibinfo {author} {\bibfnamefont
  {A.}~\bibnamefont {Aspect}},\ }\href@noop {} {\bibfield  {journal} {\bibinfo
  {journal} {Science}\ }\textbf {\bibinfo {volume} {292}},\ \bibinfo {pages}
  {461} (\bibinfo {year} {2001})}\BibitemShut {NoStop}%
\bibitem [{\citenamefont {{Pereira Dos Santos}}\ \emph
  {et~al.}(2001)\citenamefont {{Pereira Dos Santos}}, \citenamefont
  {L\'eonard}, \citenamefont {Wang}, \citenamefont {Barrelet}, \citenamefont
  {Perales}, \citenamefont {Rasel}, \citenamefont {Unnikrishnan}, \citenamefont
  {Leduc},\ and\ \citenamefont {Cohen-Tannoudji}}]{pereira2001bec}%
  \BibitemOpen
  \bibfield  {author} {\bibinfo {author} {\bibfnamefont {F.}~\bibnamefont
  {{Pereira Dos Santos}}}, \bibinfo {author} {\bibfnamefont {J.}~\bibnamefont
  {L\'eonard}}, \bibinfo {author} {\bibfnamefont {J.}~\bibnamefont {Wang}},
  \bibinfo {author} {\bibfnamefont {C.~J.}\ \bibnamefont {Barrelet}}, \bibinfo
  {author} {\bibfnamefont {F.}~\bibnamefont {Perales}}, \bibinfo {author}
  {\bibfnamefont {E.}~\bibnamefont {Rasel}}, \bibinfo {author} {\bibfnamefont
  {C.~S.}\ \bibnamefont {Unnikrishnan}}, \bibinfo {author} {\bibfnamefont
  {M.}~\bibnamefont {Leduc}}, \ and\ \bibinfo {author} {\bibfnamefont
  {C.}~\bibnamefont {Cohen-Tannoudji}},\ }\href@noop {} {\bibfield  {journal}
  {\bibinfo  {journal} {Phys.\ Rev.\ Lett.}\ }\textbf {\bibinfo {volume}
  {86}},\ \bibinfo {pages} {3459} (\bibinfo {year} {2001})}\BibitemShut
  {NoStop}%
\bibitem [{\citenamefont {McNamara}\ \emph {et~al.}(2006)\citenamefont
  {McNamara}, \citenamefont {Jeltes}, \citenamefont {Tychkov}, \citenamefont
  {Hogervorst},\ and\ \citenamefont {Vassen}}]{mcnamara2006dgb}%
  \BibitemOpen
  \bibfield  {author} {\bibinfo {author} {\bibfnamefont {J.~M.}\ \bibnamefont
  {McNamara}}, \bibinfo {author} {\bibfnamefont {T.}~\bibnamefont {Jeltes}},
  \bibinfo {author} {\bibfnamefont {A.~S.}\ \bibnamefont {Tychkov}}, \bibinfo
  {author} {\bibfnamefont {W.}~\bibnamefont {Hogervorst}}, \ and\ \bibinfo
  {author} {\bibfnamefont {W.}~\bibnamefont {Vassen}},\ }\href@noop {}
  {\bibfield  {journal} {\bibinfo  {journal} {Phys.\ Rev.\ Lett.}\ }\textbf
  {\bibinfo {volume} {97}},\ \bibinfo {pages} {080404} (\bibinfo {year}
  {2006})}\BibitemShut {NoStop}%
\bibitem [{\citenamefont {Vassen}\ \emph {et~al.}()\citenamefont {Vassen},
  \citenamefont {Cohen-Tannoudji}, \citenamefont {Leduc}, \citenamefont
  {Boiron}, \citenamefont {Westbrook}, \citenamefont {Truscott}, \citenamefont
  {Baldwin}, \citenamefont {Birkl}, \citenamefont {Cancio},\ and\ \citenamefont
  {Trippenbach}}]{vassen2011cat}%
  \BibitemOpen
  \bibfield  {author} {\bibinfo {author} {\bibfnamefont {W.}~\bibnamefont
  {Vassen}}, \bibinfo {author} {\bibfnamefont {C.}~\bibnamefont
  {Cohen-Tannoudji}}, \bibinfo {author} {\bibfnamefont {M.}~\bibnamefont
  {Leduc}}, \bibinfo {author} {\bibfnamefont {D.}~\bibnamefont {Boiron}},
  \bibinfo {author} {\bibfnamefont {C.}~\bibnamefont {Westbrook}}, \bibinfo
  {author} {\bibfnamefont {A.}~\bibnamefont {Truscott}}, \bibinfo {author}
  {\bibfnamefont {K.}~\bibnamefont {Baldwin}}, \bibinfo {author} {\bibfnamefont
  {G.}~\bibnamefont {Birkl}}, \bibinfo {author} {\bibfnamefont
  {P.}~\bibnamefont {Cancio}}, \ and\ \bibinfo {author} {\bibfnamefont
  {M.}~\bibnamefont {Trippenbach}},\ }\href@noop {} {\bibfield  {journal}
  {\bibinfo  {journal} {arXiv:1110.1361}\ }}\bibinfo {note} {(accepted for
  publication in Rev.~Mod.~Phys.)}\BibitemShut {NoStop}%
\bibitem [{\citenamefont {Jeltes}\ \emph {et~al.}(2007)\citenamefont {Jeltes},
  \citenamefont {McNamara}, \citenamefont {Hogervorst}, \citenamefont {Vassen},
  \citenamefont {Krachmalnicoff}, \citenamefont {Schellekens}, \citenamefont
  {Perrin}, \citenamefont {Chang}, \citenamefont {Boiron}, \citenamefont
  {Aspect},\ and\ \citenamefont {Westbrook}}]{jeltes2007cot}%
  \BibitemOpen
  \bibfield  {author} {\bibinfo {author} {\bibfnamefont {T.}~\bibnamefont
  {Jeltes}}, \bibinfo {author} {\bibfnamefont {J.~M.}\ \bibnamefont
  {McNamara}}, \bibinfo {author} {\bibfnamefont {W.}~\bibnamefont
  {Hogervorst}}, \bibinfo {author} {\bibfnamefont {W.}~\bibnamefont {Vassen}},
  \bibinfo {author} {\bibfnamefont {V.}~\bibnamefont {Krachmalnicoff}},
  \bibinfo {author} {\bibfnamefont {M.}~\bibnamefont {Schellekens}}, \bibinfo
  {author} {\bibfnamefont {A.}~\bibnamefont {Perrin}}, \bibinfo {author}
  {\bibfnamefont {H.}~\bibnamefont {Chang}}, \bibinfo {author} {\bibfnamefont
  {D.}~\bibnamefont {Boiron}}, \bibinfo {author} {\bibfnamefont
  {A.}~\bibnamefont {Aspect}}, \ and\ \bibinfo {author} {\bibfnamefont {C.~I.}\
  \bibnamefont {Westbrook}},\ }\href@noop {} {\bibfield  {journal} {\bibinfo
  {journal} {Nature}\ }\textbf {\bibinfo {volume} {445}},\ \bibinfo {pages}
  {402} (\bibinfo {year} {2007})}\BibitemShut {NoStop}%
\bibitem [{\citenamefont {Hodgman}\ \emph {et~al.}(2011)\citenamefont
  {Hodgman}, \citenamefont {Dall}, \citenamefont {Manning}, \citenamefont
  {Baldwin},\ and\ \citenamefont {Truscott}}]{hodgman2011dmo}%
  \BibitemOpen
  \bibfield  {author} {\bibinfo {author} {\bibfnamefont {S.~S.}\ \bibnamefont
  {Hodgman}}, \bibinfo {author} {\bibfnamefont {R.~G.}\ \bibnamefont {Dall}},
  \bibinfo {author} {\bibfnamefont {A.~G.}\ \bibnamefont {Manning}}, \bibinfo
  {author} {\bibfnamefont {K.~G.~H.}\ \bibnamefont {Baldwin}}, \ and\ \bibinfo
  {author} {\bibfnamefont {A.~G.}\ \bibnamefont {Truscott}},\ }\href@noop {}
  {\bibfield  {journal} {\bibinfo  {journal} {Science}\ }\textbf {\bibinfo
  {volume} {331}},\ \bibinfo {pages} {1046} (\bibinfo {year}
  {2011})}\BibitemShut {NoStop}%
\bibitem [{\citenamefont {Jaskula}\ \emph {et~al.}(2010)\citenamefont
  {Jaskula}, \citenamefont {Bonneau}, \citenamefont {Partridge}, \citenamefont
  {Krachmalnicoff}, \citenamefont {Deuar}, \citenamefont {Kheruntsyan},
  \citenamefont {Aspect}, \citenamefont {Boiron},\ and\ \citenamefont
  {Westbrook}}]{jaskula2010spn}%
  \BibitemOpen
  \bibfield  {author} {\bibinfo {author} {\bibfnamefont {J.~C.}\ \bibnamefont
  {Jaskula}}, \bibinfo {author} {\bibfnamefont {M.}~\bibnamefont {Bonneau}},
  \bibinfo {author} {\bibfnamefont {G.~B.}\ \bibnamefont {Partridge}}, \bibinfo
  {author} {\bibfnamefont {V.}~\bibnamefont {Krachmalnicoff}}, \bibinfo
  {author} {\bibfnamefont {P.}~\bibnamefont {Deuar}}, \bibinfo {author}
  {\bibfnamefont {K.~V.}\ \bibnamefont {Kheruntsyan}}, \bibinfo {author}
  {\bibfnamefont {A.}~\bibnamefont {Aspect}}, \bibinfo {author} {\bibfnamefont
  {D.}~\bibnamefont {Boiron}}, \ and\ \bibinfo {author} {\bibfnamefont {C.~I.}\
  \bibnamefont {Westbrook}},\ }\href@noop {} {\bibfield  {journal} {\bibinfo
  {journal} {Phys.\ Rev.\ Lett.}\ }\textbf {\bibinfo {volume} {105}},\ \bibinfo
  {pages} {190402} (\bibinfo {year} {2010})}\BibitemShut {NoStop}%
\bibitem [{\citenamefont {{van Rooij}}\ \emph {et~al.}(2011)\citenamefont {{van
  Rooij}}, \citenamefont {Borbely}, \citenamefont {Simonet}, \citenamefont
  {Hoogerland}, \citenamefont {Eikema}, \citenamefont {Rozendaal},\ and\
  \citenamefont {Vassen}}]{rooij2011fmi}%
  \BibitemOpen
  \bibfield  {author} {\bibinfo {author} {\bibfnamefont {R.}~\bibnamefont {{van
  Rooij}}}, \bibinfo {author} {\bibfnamefont {J.~S.}\ \bibnamefont {Borbely}},
  \bibinfo {author} {\bibfnamefont {J.}~\bibnamefont {Simonet}}, \bibinfo
  {author} {\bibfnamefont {M.~D.}\ \bibnamefont {Hoogerland}}, \bibinfo
  {author} {\bibfnamefont {K.~S.~E.}\ \bibnamefont {Eikema}}, \bibinfo {author}
  {\bibfnamefont {R.~A.}\ \bibnamefont {Rozendaal}}, \ and\ \bibinfo {author}
  {\bibfnamefont {W.}~\bibnamefont {Vassen}},\ }\href@noop {} {\bibfield
  {journal} {\bibinfo  {journal} {Science}\ }\textbf {\bibinfo {volume}
  {333}},\ \bibinfo {pages} {196} (\bibinfo {year} {2011})}\BibitemShut
  {NoStop}%
\bibitem [{\citenamefont {Julienne}\ and\ \citenamefont
  {Mies}(1989)}]{julienne1989cou}%
  \BibitemOpen
  \bibfield  {author} {\bibinfo {author} {\bibfnamefont {P.~S.}\ \bibnamefont
  {Julienne}}\ and\ \bibinfo {author} {\bibfnamefont {F.~H.}\ \bibnamefont
  {Mies}},\ }\href@noop {} {\bibfield  {journal} {\bibinfo  {journal} {J.\
  Opt.\ Soc.\ Am.\ B}\ }\textbf {\bibinfo {volume} {6}},\ \bibinfo {pages}
  {2257} (\bibinfo {year} {1989})}\BibitemShut {NoStop}%
\bibitem [{\citenamefont {Shlyapnikov}\ \emph {et~al.}(1994)\citenamefont
  {Shlyapnikov}, \citenamefont {Walraven}, \citenamefont {Rahmanov},\ and\
  \citenamefont {Reynolds}}]{shlyapnikov1994dka}%
  \BibitemOpen
  \bibfield  {author} {\bibinfo {author} {\bibfnamefont {G.~V.}\ \bibnamefont
  {Shlyapnikov}}, \bibinfo {author} {\bibfnamefont {J.~T.~M.}\ \bibnamefont
  {Walraven}}, \bibinfo {author} {\bibfnamefont {U.~M.}\ \bibnamefont
  {Rahmanov}}, \ and\ \bibinfo {author} {\bibfnamefont {M.~W.}\ \bibnamefont
  {Reynolds}},\ }\href@noop {} {\bibfield  {journal} {\bibinfo  {journal}
  {Phys.\ Rev.\ Lett.}\ }\textbf {\bibinfo {volume} {73}},\ \bibinfo {pages}
  {3247} (\bibinfo {year} {1994})}\BibitemShut {NoStop}%
\bibitem [{\citenamefont {Fedichev}\ \emph
  {et~al.}(1996{\natexlab{a}})\citenamefont {Fedichev}, \citenamefont
  {Reynolds}, \citenamefont {Rahmanov},\ and\ \citenamefont
  {Shlyapnikov}}]{fedichev1996idp}%
  \BibitemOpen
  \bibfield  {author} {\bibinfo {author} {\bibfnamefont {P.~O.}\ \bibnamefont
  {Fedichev}}, \bibinfo {author} {\bibfnamefont {M.~W.}\ \bibnamefont
  {Reynolds}}, \bibinfo {author} {\bibfnamefont {U.~M.}\ \bibnamefont
  {Rahmanov}}, \ and\ \bibinfo {author} {\bibfnamefont {G.~V.}\ \bibnamefont
  {Shlyapnikov}},\ }\href@noop {} {\bibfield  {journal} {\bibinfo  {journal}
  {Phys.\ Rev.\ A}\ }\textbf {\bibinfo {volume} {53}},\ \bibinfo {pages} {1447}
  (\bibinfo {year} {1996}{\natexlab{a}})}\BibitemShut {NoStop}%
\bibitem [{\citenamefont {Venturi}\ \emph {et~al.}(1999)\citenamefont
  {Venturi}, \citenamefont {Whittingham}, \citenamefont {Leo},\ and\
  \citenamefont {Peach}}]{venturi1999ccc}%
  \BibitemOpen
  \bibfield  {author} {\bibinfo {author} {\bibfnamefont {V.}~\bibnamefont
  {Venturi}}, \bibinfo {author} {\bibfnamefont {I.~B.}\ \bibnamefont
  {Whittingham}}, \bibinfo {author} {\bibfnamefont {P.~J.}\ \bibnamefont
  {Leo}}, \ and\ \bibinfo {author} {\bibfnamefont {G.}~\bibnamefont {Peach}},\
  }\href@noop {} {\bibfield  {journal} {\bibinfo  {journal} {Phys.\ Rev.\ A}\
  }\textbf {\bibinfo {volume} {60}},\ \bibinfo {pages} {4635} (\bibinfo {year}
  {1999})}\BibitemShut {NoStop}%
\bibitem [{\citenamefont {Leo}\ \emph {et~al.}(2001)\citenamefont {Leo},
  \citenamefont {Venturi}, \citenamefont {Whittingham},\ and\ \citenamefont
  {Babb}}]{leo2001uco}%
  \BibitemOpen
  \bibfield  {author} {\bibinfo {author} {\bibfnamefont {P.~J.}\ \bibnamefont
  {Leo}}, \bibinfo {author} {\bibfnamefont {V.}~\bibnamefont {Venturi}},
  \bibinfo {author} {\bibfnamefont {I.~B.}\ \bibnamefont {Whittingham}}, \ and\
  \bibinfo {author} {\bibfnamefont {J.~F.}\ \bibnamefont {Babb}},\ }\href@noop
  {} {\bibfield  {journal} {\bibinfo  {journal} {Phys.\ Rev.\ A}\ }\textbf
  {\bibinfo {volume} {64}},\ \bibinfo {pages} {042710} (\bibinfo {year}
  {2001})}\BibitemShut {NoStop}%
\bibitem [{\citenamefont {Herschbach}\ \emph {et~al.}(2000)\citenamefont
  {Herschbach}, \citenamefont {Tol}, \citenamefont {Hogervorst},\ and\
  \citenamefont {Vassen}}]{herschbach2000sop}%
  \BibitemOpen
  \bibfield  {author} {\bibinfo {author} {\bibfnamefont {N.}~\bibnamefont
  {Herschbach}}, \bibinfo {author} {\bibfnamefont {P.~J.~J.}\ \bibnamefont
  {Tol}}, \bibinfo {author} {\bibfnamefont {W.}~\bibnamefont {Hogervorst}}, \
  and\ \bibinfo {author} {\bibfnamefont {W.}~\bibnamefont {Vassen}},\
  }\href@noop {} {\bibfield  {journal} {\bibinfo  {journal} {Phys. Rev. A}\
  }\textbf {\bibinfo {volume} {61}},\ \bibinfo {pages} {050702} (\bibinfo
  {year} {2000})}\BibitemShut {NoStop}%
\bibitem [{\citenamefont {Sirjean}\ \emph {et~al.}(2002)\citenamefont
  {Sirjean}, \citenamefont {Seidelin}, \citenamefont {Gomes}, \citenamefont
  {Boiron}, \citenamefont {Westbrook}, \citenamefont {Aspect},\ and\
  \citenamefont {Shlyapnikov}}]{sirjean2002iri}%
  \BibitemOpen
  \bibfield  {author} {\bibinfo {author} {\bibfnamefont {O.}~\bibnamefont
  {Sirjean}}, \bibinfo {author} {\bibfnamefont {S.}~\bibnamefont {Seidelin}},
  \bibinfo {author} {\bibfnamefont {J.~V.}\ \bibnamefont {Gomes}}, \bibinfo
  {author} {\bibfnamefont {D.}~\bibnamefont {Boiron}}, \bibinfo {author}
  {\bibfnamefont {C.~I.}\ \bibnamefont {Westbrook}}, \bibinfo {author}
  {\bibfnamefont {A.}~\bibnamefont {Aspect}}, \ and\ \bibinfo {author}
  {\bibfnamefont {G.~V.}\ \bibnamefont {Shlyapnikov}},\ }\href@noop {}
  {\bibfield  {journal} {\bibinfo  {journal} {Phys.\ Rev.\ Lett.}\ }\textbf
  {\bibinfo {volume} {89}},\ \bibinfo {pages} {220406} (\bibinfo {year}
  {2002})}\BibitemShut {NoStop}%
\bibitem [{\citenamefont {Seidelin}\ \emph {et~al.}(2004)\citenamefont
  {Seidelin}, \citenamefont {{Viana Gomes}}, \citenamefont {Hoppeler},
  \citenamefont {Sirjean}, \citenamefont {Boiron}, \citenamefont {Aspect},\
  and\ \citenamefont {Westbrook}}]{seidelin2004gte}%
  \BibitemOpen
  \bibfield  {author} {\bibinfo {author} {\bibfnamefont {S.}~\bibnamefont
  {Seidelin}}, \bibinfo {author} {\bibfnamefont {J.~V.}~\bibnamefont {{Gomes}}}, 
  \bibinfo {author} {\bibfnamefont {R.}~\bibnamefont {Hoppeler}},
  \bibinfo {author} {\bibfnamefont {O.}~\bibnamefont {Sirjean}}, \bibinfo
  {author} {\bibfnamefont {D.}~\bibnamefont {Boiron}}, \bibinfo {author}
  {\bibfnamefont {A.}~\bibnamefont {Aspect}}, \ and\ \bibinfo {author}
  {\bibfnamefont {C.~I.}\ \bibnamefont {Westbrook}},\ }\href@noop {} {\bibfield
   {journal} {\bibinfo  {journal} {Phys.\ Rev.\ Lett.}\ }\textbf {\bibinfo
  {volume} {93}},\ \bibinfo {pages} {090409} (\bibinfo {year}
  {2004})}\BibitemShut {NoStop}%
\bibitem [{\citenamefont {Tychkov}\ \emph {et~al.}(2006)\citenamefont
  {Tychkov}, \citenamefont {Jeltes}, \citenamefont {McNamara}, \citenamefont
  {Tol}, \citenamefont {Herschbach}, \citenamefont {Hogervorst},\ and\
  \citenamefont {Vassen}}]{tychkov2006mtb}%
  \BibitemOpen
  \bibfield  {author} {\bibinfo {author} {\bibfnamefont {A.~S.}\ \bibnamefont
  {Tychkov}}, \bibinfo {author} {\bibfnamefont {T.}\ \bibnamefont {Jeltes}},
  \bibinfo {author} {\bibfnamefont {J.~M.}\ \bibnamefont {McNamara}}, \bibinfo
  {author} {\bibfnamefont {P.~J.~J.}\ \bibnamefont {Tol}}, \bibinfo {author}
  {\bibfnamefont {N.}~\bibnamefont {Herschbach}}, \bibinfo {author}
  {\bibfnamefont {W.}~\bibnamefont {Hogervorst}}, \ and\ \bibinfo {author}
  {\bibfnamefont {W.}~\bibnamefont {Vassen}},\ }\href@noop {} {\bibfield
  {journal} {\bibinfo  {journal} {Phys.\ Rev.\ A}\ }\textbf {\bibinfo {volume}
  {73}},\ \bibinfo {pages} {031603(R)} (\bibinfo {year} {2006})}\BibitemShut
  {NoStop}%
\bibitem [{\citenamefont {Fedichev}\ \emph
  {et~al.}(1996{\natexlab{b}})\citenamefont {Fedichev}, \citenamefont
  {Reynolds},\ and\ \citenamefont {Shlyapnikov}}]{fedichev1996tbr}%
  \BibitemOpen
  \bibfield  {author} {\bibinfo {author} {\bibfnamefont {P.~O.}\ \bibnamefont
  {Fedichev}}, \bibinfo {author} {\bibfnamefont {M.~W.}\ \bibnamefont
  {Reynolds}}, \ and\ \bibinfo {author} {\bibfnamefont {G.~V.}\ \bibnamefont
  {Shlyapnikov}},\ }\href@noop {} {\bibfield  {journal} {\bibinfo  {journal}
  {Phys. Rev. Lett.}\ }\textbf {\bibinfo {volume} {77}},\ \bibinfo {pages}
  {2921} (\bibinfo {year} {1996}{\natexlab{b}})}\BibitemShut {NoStop}%
\bibitem [{\citenamefont {Partridge}\ \emph {et~al.}(2010)\citenamefont
  {Partridge}, \citenamefont {Jaskula}, \citenamefont {Bonneau}, \citenamefont
  {Boiron},\ and\ \citenamefont {Westbrook}}]{partridge2010bec}%
  \BibitemOpen
  \bibfield  {author} {\bibinfo {author} {\bibfnamefont {G.~B.}\ \bibnamefont
  {Partridge}}, \bibinfo {author} {\bibfnamefont {J.-C.}\ \bibnamefont
  {Jaskula}}, \bibinfo {author} {\bibfnamefont {M.}~\bibnamefont {Bonneau}},
  \bibinfo {author} {\bibfnamefont {D.}~\bibnamefont {Boiron}}, \ and\ \bibinfo
  {author} {\bibfnamefont {C.~I.}\ \bibnamefont {Westbrook}},\ }\href@noop {}
  {\bibfield  {journal} {\bibinfo  {journal} {Phys.\ Rev.\ A}\ }\textbf
  {\bibinfo {volume} {81}},\ \bibinfo {pages} {053631} (\bibinfo {year}
  {2010})}\BibitemShut {NoStop}%
\bibitem [{\citenamefont {Dall}\ \emph {et~al.}(2011)\citenamefont {Dall},
  \citenamefont {Hodgman}, \citenamefont {Manning}, \citenamefont {Johnsson},
  \citenamefont {Baldwin},\ and\ \citenamefont {Truscott}}]{dall2011ooa}%
  \BibitemOpen
  \bibfield  {author} {\bibinfo {author} {\bibfnamefont {R.~G.}\ \bibnamefont
  {Dall}}, \bibinfo {author} {\bibfnamefont {S.~S.}\ \bibnamefont {Hodgman}},
  \bibinfo {author} {\bibfnamefont {A.~G.}\ \bibnamefont {Manning}}, \bibinfo
  {author} {\bibfnamefont {M.~T.}\ \bibnamefont {Johnsson}}, \bibinfo {author}
  {\bibfnamefont {K.~G.~H.}\ \bibnamefont {Baldwin}}, \ and\ \bibinfo {author}
  {\bibfnamefont {A.~G.}\ \bibnamefont {Truscott}},\ }\href@noop {} {\bibfield
  {journal} {\bibinfo  {journal} {Nature Comm.}\ }\textbf {\bibinfo {volume}
  {2}},\ \bibinfo {pages} {291} (\bibinfo {year} {2011})}\BibitemShut {NoStop}%
\bibitem [{\citenamefont {Goosen}\ \emph {et~al.}(2010)\citenamefont {Goosen},
  \citenamefont {Tiecke}, \citenamefont {Vassen},\ and\ \citenamefont
  {Kokkelmans}}]{goosen2010fri}%
  \BibitemOpen
  \bibfield  {author} {\bibinfo {author} {\bibfnamefont {M.~R.}\ \bibnamefont
  {Goosen}}, \bibinfo {author} {\bibfnamefont {T.~G.}\ \bibnamefont {Tiecke}},
  \bibinfo {author} {\bibfnamefont {W.}~\bibnamefont {Vassen}}, \ and\ \bibinfo
  {author} {\bibfnamefont {S.~J.~J.~M.~F.}\ \bibnamefont {Kokkelmans}},\
  }\href@noop {} {\bibfield  {journal} {\bibinfo  {journal} {Phys.\ Rev.\ A}\
  }\textbf {\bibinfo {volume} {82}},\ \bibinfo {pages} {042713} (\bibinfo
  {year} {2010})}\BibitemShut {NoStop}%
\bibitem [{\citenamefont {Moal}\ \emph {et~al.}(2006)\citenamefont {Moal},
  \citenamefont {Portier}, \citenamefont {Kim}, \citenamefont {Dugu\'e},
  \citenamefont {Rapol}, \citenamefont {Leduc},\ and\ \citenamefont
  {Cohen-Tannoudji}}]{moal2006ado}%
  \BibitemOpen
  \bibfield  {author} {\bibinfo {author} {\bibfnamefont {S.}~\bibnamefont
  {Moal}}, \bibinfo {author} {\bibfnamefont {M.}~\bibnamefont {Portier}},
  \bibinfo {author} {\bibfnamefont {J.}~\bibnamefont {Kim}}, \bibinfo {author}
  {\bibfnamefont {J.}~\bibnamefont {Dugu\'e}}, \bibinfo {author} {\bibfnamefont
  {U.~D.}\ \bibnamefont {Rapol}}, \bibinfo {author} {\bibfnamefont
  {M.}~\bibnamefont {Leduc}}, \ and\ \bibinfo {author} {\bibfnamefont
  {C.}~\bibnamefont {Cohen-Tannoudji}},\ }\href@noop {} {\bibfield  {journal}
  {\bibinfo  {journal} {Phys.\ Rev.\ Lett.}\ }\textbf {\bibinfo {volume}
  {96}},\ \bibinfo {pages} {023203} (\bibinfo {year} {2006})}\BibitemShut
  {NoStop}%
\bibitem [{\citenamefont {St\"arck}\ and\ \citenamefont
  {Meyer}(1994)}]{stark1994lri}%
  \BibitemOpen
  \bibfield  {author} {\bibinfo {author} {\bibfnamefont {J.}~\bibnamefont
  {St\"arck}}\ and\ \bibinfo {author} {\bibfnamefont {W.}~\bibnamefont
  {Meyer}},\ }\href@noop {} {\bibfield  {journal} {\bibinfo  {journal} {Chem.\
  Phys.\ Lett.}\ }\textbf {\bibinfo {volume} {225}},\ \bibinfo {pages} {229}
  (\bibinfo {year} {1994})}\BibitemShut {NoStop}%
\bibitem [{\citenamefont {M\"uller}\ \emph {et~al.}(1991)\citenamefont
  {M\"uller}, \citenamefont {Merz}, \citenamefont {Ruf}, \citenamefont {Hotop},
  \citenamefont {Meyer},\ and\ \citenamefont {Movre}}]{muller1991eat}%
  \BibitemOpen
  \bibfield  {author} {\bibinfo {author} {\bibfnamefont {M.~W.}\ \bibnamefont
  {M\"uller}}, \bibinfo {author} {\bibfnamefont {A.}~\bibnamefont {Merz}},
  \bibinfo {author} {\bibfnamefont {M.-W.}\ \bibnamefont {Ruf}}, \bibinfo
  {author} {\bibfnamefont {H.}~\bibnamefont {Hotop}}, \bibinfo {author}
  {\bibfnamefont {W.}~\bibnamefont {Meyer}}, \ and\ \bibinfo {author}
  {\bibfnamefont {M.}~\bibnamefont {Movre}},\ }\href@noop {} {\bibfield
  {journal} {\bibinfo  {journal} {Z.\ Phys.\ D}\ }\textbf {\bibinfo {volume}
  {21}},\ \bibinfo {pages} {89} (\bibinfo {year} {1991})}\BibitemShut {NoStop}%
\bibitem [{\citenamefont {Przybytek}\ and\ \citenamefont
  {Jeziorski}(2005)}]{przybytek2005bft}%
  \BibitemOpen
  \bibfield  {author} {\bibinfo {author} {\bibfnamefont {M.}~\bibnamefont
  {Przybytek}}\ and\ \bibinfo {author} {\bibfnamefont {B.}~\bibnamefont
  {Jeziorski}},\ }\href@noop {} {\bibfield  {journal} {\bibinfo  {journal} {J.\
  Chem.\ Phys.}\ }\textbf {\bibinfo {volume} {123}},\ \bibinfo {pages} {134315}
  (\bibinfo {year} {2005})}\BibitemShut {NoStop}%
\bibitem [{\citenamefont {Simonet}(2011)}]{simonet2011PhD}%
  \BibitemOpen
  \bibfield  {author} {\bibinfo {author} {\bibfnamefont {J.}~\bibnamefont
  {Simonet}},\ }\emph {\bibinfo {title} {Optical traps for Ultracold Metastable
  Helium atoms}},\ \href@noop {} {\bibinfo {type} {{PhD} thesis}},\ \bibinfo
  {school} {$\acute{\rm E}$cole Normale Sup$\acute{\rm e}$rieure, Labortoire
  Kastler-Brossel} (\bibinfo {year} {2011})\BibitemShut {NoStop}%
\bibitem [{\citenamefont {Braaten}\ and\ \citenamefont
  {Hammer}(2006)}]{braaten2006uif}%
  \BibitemOpen
  \bibfield  {author} {\bibinfo {author} {\bibfnamefont {E.}~\bibnamefont
  {Braaten}}\ and\ \bibinfo {author} {\bibfnamefont {H.-W.}\ \bibnamefont
  {Hammer}},\ }\href@noop {} {\bibfield  {journal} {\bibinfo  {journal} {Phys.
  Rep.}\ }\textbf {\bibinfo {volume} {428}},\ \bibinfo {pages} {259} (\bibinfo
  {year} {2006})}\BibitemShut {NoStop}%
\bibitem [{\citenamefont {Weber}\ \emph {et~al.}(2003)\citenamefont {Weber},
  \citenamefont {Herbig}, \citenamefont {Mark}, \citenamefont {N\"agerl},\ and\
  \citenamefont {Grimm}}]{weber2003tbr}%
  \BibitemOpen
  \bibfield  {author} {\bibinfo {author} {\bibfnamefont {T.}~\bibnamefont
  {Weber}}, \bibinfo {author} {\bibfnamefont {J.}~\bibnamefont {Herbig}},
  \bibinfo {author} {\bibfnamefont {M.}~\bibnamefont {Mark}}, \bibinfo {author}
  {\bibfnamefont {H.-C.}\ \bibnamefont {N\"agerl}}, \ and\ \bibinfo {author}
  {\bibfnamefont {R.}~\bibnamefont {Grimm}},\ }\href@noop {} {\bibfield
  {journal} {\bibinfo  {journal} {Phys.\ Rev.\ Lett.}\ }\textbf {\bibinfo
  {volume} {91}},\ \bibinfo {pages} {123201} (\bibinfo {year}
  {2003})}\BibitemShut {NoStop}%
\bibitem [{Note1()}]{Note1}%
  \BibitemOpen
  \bibinfo {note} {The van der Waals length $r_{\protect \text {vdW}}=\protect
  \frac {1}{2}(m C_6/\hbar ^2)^{1/4}$=34$a_0$, with $C_6$=3276.680 a.\protect
  \tmspace +\thinmuskip {.1667em}u.\ from~\cite {przybytek2005bft}}\BibitemShut
  {NoStop}%
\bibitem [{\citenamefont {Kagan}\ \emph {et~al.}(1985)\citenamefont {Kagan},
  \citenamefont {Svistunov},\ and\ \citenamefont {Shlyapnikov}}]{kagan1985}%
  \BibitemOpen
  \bibfield  {author} {\bibinfo {author} {\bibfnamefont {Y.}~\bibnamefont
  {Kagan}}, \bibinfo {author} {\bibfnamefont {B.~V.}\ \bibnamefont
  {Svistunov}}, \ and\ \bibinfo {author} {\bibfnamefont {G.~V.}\ \bibnamefont
  {Shlyapnikov}},\ }\href@noop {} {\bibfield  {journal} {\bibinfo  {journal}
  {JETP Lett.}\ }\textbf {\bibinfo {volume} {42}},\ \bibinfo {pages} {209}
  (\bibinfo {year} {1985})}\BibitemShut {NoStop}%
\bibitem [{\citenamefont {Pitaevskii}\ and\ \citenamefont
  {Stringari}(2003)}]{pitaevskii2003book}%
  \BibitemOpen
  \bibfield  {author} {\bibinfo {author} {\bibfnamefont {L.}~\bibnamefont
  {Pitaevskii}}\ and\ \bibinfo {author} {\bibfnamefont {S.}~\bibnamefont
  {Stringari}},\ }\href@noop {} {\emph {\bibinfo {title} {Bose-Einstein
  Condensation}}}\ (\bibinfo  {publisher} {Oxford Science Publications, New
  York},\ \bibinfo {year} {2003})\BibitemShut {NoStop}%
\bibitem [{\citenamefont {Knoop}\ \emph {et~al.}(2011)\citenamefont {Knoop},
  \citenamefont {Schuster}, \citenamefont {Scelle}, \citenamefont {Trautmann},
  \citenamefont {Appmeier}, \citenamefont {Oberthaler}, \citenamefont
  {Tiesinga},\ and\ \citenamefont {Tiemann}}]{knoop2011fsa}%
  \BibitemOpen
  \bibfield  {author} {\bibinfo {author} {\bibfnamefont {S.}~\bibnamefont
  {Knoop}}, \bibinfo {author} {\bibfnamefont {T.}~\bibnamefont {Schuster}},
  \bibinfo {author} {\bibfnamefont {R.}~\bibnamefont {Scelle}}, \bibinfo
  {author} {\bibfnamefont {A.}~\bibnamefont {Trautmann}}, \bibinfo {author}
  {\bibfnamefont {J.}~\bibnamefont {Appmeier}}, \bibinfo {author}
  {\bibfnamefont {M.~K.}\ \bibnamefont {Oberthaler}}, \bibinfo {author}
  {\bibfnamefont {E.}~\bibnamefont {Tiesinga}}, \ and\ \bibinfo {author}
  {\bibfnamefont {E.}~\bibnamefont {Tiemann}},\ }\href@noop {} {\bibfield
  {journal} {\bibinfo  {journal} {Phys.\ Rev.\ A}\ }\textbf {\bibinfo {volume}
  {83}},\ \bibinfo {pages} {042704} (\bibinfo {year} {2011})}\BibitemShut
  {NoStop}%
\bibitem [{\citenamefont {Hammer}\ \emph {et~al.}(2007)\citenamefont {Hammer},
  \citenamefont {L\"ahde},\ and\ \citenamefont {Platter}}]{hammer2007erc}%
  \BibitemOpen
  \bibfield  {author} {\bibinfo {author} {\bibfnamefont {H.-W.}\ \bibnamefont
  {Hammer}}, \bibinfo {author} {\bibfnamefont {T.~A.}\ \bibnamefont {L\"ahde}},
  \ and\ \bibinfo {author} {\bibfnamefont {L.}~\bibnamefont {Platter}},\
  }\href@noop {} {\bibfield  {journal} {\bibinfo  {journal} {Phys.\ Rev.\ A}\
  }\textbf {\bibinfo {volume} {75}},\ \bibinfo {pages} {032715} (\bibinfo
  {year} {2007})}\BibitemShut {NoStop}%
\bibitem [{\citenamefont {Platter}\ \emph {et~al.}(2009)\citenamefont
  {Platter}, \citenamefont {Ji},\ and\ \citenamefont
  {Phillips}}]{platter2009rct}%
  \BibitemOpen
  \bibfield  {author} {\bibinfo {author} {\bibfnamefont {L.}~\bibnamefont
  {Platter}}, \bibinfo {author} {\bibfnamefont {C.}~\bibnamefont {Ji}}, \ and\
  \bibinfo {author} {\bibfnamefont {D.~R.}\ \bibnamefont {Phillips}},\
  }\href@noop {} {\bibfield  {journal} {\bibinfo  {journal} {Phys.\ Rev.\ A}\
  }\textbf {\bibinfo {volume} {79}},\ \bibinfo {pages} {022702} (\bibinfo
  {year} {2009})}\BibitemShut {NoStop}%
\bibitem [{\citenamefont {Ji}\ \emph {et~al.}(2010)\citenamefont {Ji},
  \citenamefont {Phillips},\ and\ \citenamefont {Platter}}]{ji2010bui}%
  \BibitemOpen
  \bibfield  {author} {\bibinfo {author} {\bibfnamefont {C.}~\bibnamefont
  {Ji}}, \bibinfo {author} {\bibfnamefont {D.~R.}\ \bibnamefont {Phillips}}, \
  and\ \bibinfo {author} {\bibfnamefont {L.}~\bibnamefont {Platter}},\
  }\href@noop {} {\bibfield  {journal} {\bibinfo  {journal} {Europhys.\ Lett.}\
  }\textbf {\bibinfo {volume} {92}},\ \bibinfo {pages} {13003} (\bibinfo {year}
  {2010})}\BibitemShut {NoStop}%
\end{thebibliography}
\end{document}